\documentclass[12pt]{article}
\usepackage{amsfonts,amssymb}

\def\C{\mathbb{C}}
\def\N{\mathbb{N}}
\def\Z{\mathbb{Z}}
\def\D{\mathbb{D}}
\def\R{\mathbb{R}}

\def\bq{ \begin{equation} }
\def\eq{ \end{equation} }
\def\ben{ \begin{eqnarray} }
\def\en{ \end{eqnarray} }

\def\frac#1#2{{#1\over #2}}

\def\on#1#2{\mathop{\vbox{\ialign{##\crcr\noalign{\kern2pt}
$\scriptstyle{#2}$\crcr\noalign{\kern2pt\nointerlineskip}
\kern-2pt$\hfil\displaystyle{#1}\hfil$\crcr}}}\limits}

\begin{document}

\title{Integrable
Lagrangians and modular forms}
\author{E.V. Ferapontov and  A.V. Odesskii}
   \date{}
\vspace{-20mm}
   \maketitle
\vspace{-7mm}
\begin{center}
Department of Mathematical Sciences \\
Loughborough University \\
Loughborough, Leicestershire LE11 3TU, UK \\[1ex]
and \\[1ex]
School of Mathematics\\
University of Manchester \\
Oxford Road, Manchester M13 9PL, UK\\[1ex]
e-mails: \\
\texttt{E.V.Ferapontov@lboro.ac.uk}\\
\texttt{Alexander.Odesskii@manchester.ac.uk}
\end{center}

\medskip

\begin{abstract}
We  investigate  non-degenerate Lagrangians of the form
 $$
 \int f(u_x, u_y, u_t) \, dx\, dy\, dt
 $$
such that the corresponding Euler-Lagrange equations
 $
 (f_{u_x})_x+ (f_{u_y})_y+ (f_{u_t})_t=0
 $
are integrable by the method of hydrodynamic reductions. We
demonstrate  that the integrability conditions, which constitute an
involutive over-determined system of fourth order PDEs for the
Lagrangian density $f$, are  invariant under a $20$-parameter group
of Lie-point symmetries whose action on the moduli space of
integrable Lagrangians has an open orbit. The density of the
`master-Lagrangian'   corresponding to this orbit is shown to be a
modular form in three variables defined on a complex hyperbolic
ball. We demonstrate how the knowledge of the symmetry group  allows
one to linearise the integrability conditions.

\medskip


\medskip

\end{abstract}

\newpage

 \section{Introduction}

In this paper we investigate integrable
three-dimensional Euler-Lagrange equations,
\begin{equation}
 (f_{u_x})_x+ (f_{u_y})_y+ (f_{u_t})_t=0,
\label{Lag}
\end{equation}
 corresponding to  Lagrangian densities of the form $ f(u_x, u_y, u_t)$.  Familiar examples include  the dispersionless KP equation $u_{xt}-u_xu_{xx}=u_{yy}$ with the Lagrangian density  $f=\frac{1}{3}u_x^3+u_y^2-u_xu_t$; this equation, also known as the Khokhlov-Zabolotskaya equation, arises in non-linear acoustics \cite{KZ}. Another example,  $u_{xx}+u_{yy}=e^{u_t}u_{tt}$,  is known as the Boyer-Finley equation \cite{BF}:  it  appears as a symmetry reduction of the self-duality equations, and corresponds to the  Lagrangian density $f=u_x^2+u_y^2-2e^{u_t}$.

  The paper \cite{FKT} provides a system of partial differential equations for the Lagrangian
density $f(a, b, c)$ (we  set $a=u_x, \ b=u_y, \ c=u_t$)
which are  necessary and sufficient  for the integrability of the equation (\ref{Lag}) by the method of hydrodynamic reductions as proposed in \cite{F1}. These conditions  can be
represented in a remarkable compact form:

{\bf Theorem 1}  \cite{FKT}. {\it For a  non-degenerate Lagrangian,
the Euler-Lagrange equation (\ref{Lag}) is integrable by the method
of hydrodynamic reductions if and only if the density $f$ satisfies
the relation}
\begin{equation}
d^4f=d^3f\frac{dH}{H}+\frac{3}{H}{ det} (dM);
\label{fourth}
\end{equation}
here $d^3f$ and $d^4f$ are the symmetric differentials of $f$. The Hessian $H$ and the $4 \times
4$ matrix $M$ are defined as follows:
\begin{equation}
H=det
\left(\begin{array}{ccc}
f_{aa} & f_{ab} & f_{ac} \\
f_{ab} & f_{bb} & f_{bc} \\
f_{ac} & f_{bc} & f_{cc}
\end{array}
\right), ~~~
M=\left(\begin{array}{cccc}
0 & f_a & f_b & f_c \\
f_a & f_{aa} & f_{ab} & f_{ac} \\
f_b & f_{ab} & f_{bb} & f_{bc} \\
f_c & f_{ac} & f_{bc} & f_{cc}
\end{array}
\right).
\label{Hessian}
\end{equation}
The differential $dM=M_ada+M_bdb+M_cdc$ is a matrix-valued   form
$$
\left(\begin{array}{cccc}
0 & f_{aa} & f_{ab} & f_{ac} \\
f_{aa} & f_{aaa} & f_{aab} & f_{aac} \\
f_{ab} & f_{aab} & f_{abb} & f_{abc} \\
f_{ac} & f_{aac} & f_{abc} & f_{acc}
\end{array}
\right)da+
\left(\begin{array}{cccc}
0 & f_{ab} & f_{bb} & f_{bc} \\
f_{ab} & f_{aab} & f_{abb} & f_{abc} \\
f_{bb} & f_{abb} & f_{bbb} & f_{bbc} \\
f_{bc} & f_{abc} & f_{bbc} & f_{bcc}
\end{array}
\right)db+
\left(\begin{array}{cccc}
0 & f_{ac} & f_{bc} & f_{cc} \\
f_{ac} & f_{aac} & f_{abc} & f_{acc} \\
f_{bc} & f_{abc} & f_{bbc} & f_{bcc} \\
f_{cc} & f_{acc} & f_{bcc} & f_{ccc}
\end{array}
\right)dc.
$$
A Lagrangian is said to be non-degenerate iff $H \ne 0$ (we point out that the equations $H=0$  and $det M=0$ have been discussed in the literature, see \cite{Fa} and references therein).

Both sides of the relation (\ref{fourth}) are homogeneous symmetric quartics
 in  $da, db, dc$. Equating similar terms we obtain expressions for {\it all}  fourth order partial
derivatives of the density $f$ in terms of its second and third order derivatives (15 equations altogether). The resulting over-determined system for $f$ is in involution, and its solution
space is $20$-dimensional: indeed, the values of partial derivatives of $f$ up to
order 3 at a  point $(a_0, b_0, c_0)$  amount to  $20$ arbitrary constants. Thus, we are dealing with a $20$-dimensional moduli space of integrable Lagrangians.

In Sect. 2 we  prove  that the integrability conditions (\ref{fourth}) are invariant under a $20$-parameter
group of Lie-point symmetries whose action on the  moduli space of integrable Lagrangians possesses an open orbit.

Explicit formulae for integrable Lagrangians in terms of modular forms are constructed in Sect. 3. We first consider  Lagrangian densities of the form $f=u_x u_yg(u_t)$, which can be viewed as a deformation of the integrable  density $f=u_x u_y u_t$ found in
\cite{FKT}. By virtue of the integrability conditions (\ref{fourth}), the function $g$ has to satisfy the fourth order ODE
$$
g''''(g^2g''-2g(g')^2)-9(g')^2(g'')^2+2gg'g''g'''+8(g')^3g'''-g^2(g''')^2=0,
$$
which inherits a remarkable $Gl(2, \R)$-invariance. We prove that he `generic' solution of this ODE is given by the series
$$
g(u_t)=\sum_{(\alpha,\beta)\in\Z^2}e^{(\alpha^2-\alpha\beta+\beta^2)u_t}=1+6e^{u_t}+6e^{3u_t}+6e^{4u_t}+12e^{7u_t}+....;
$$
notice that under the substitution $u_t=2\pi i z$ the right hand side of this formula becomes a special modular form of weight one and level three, known as the Eisenstein series $E_{1, 3}(z)$. We point out that modular forms and non-linear ODEs related to them appear in a variety of problems  in mathematical physics, see e.g.  \cite {A1, A2, A3, Clarkson, H1, H2, O1, O2, Takh} and references therein.

Lagrangian densities of the  form  $g(u_x, u_y) u_t$ and the general case $f(u_x, u_y, u_t)$ are discussed in Sect. 3.2 and 3.3, respectively.  Here the `generic'  solution is an automorphic  form of  two (three) variables.

 \section{Symmetry group of the problem}

The first main observation, overlooked in \cite{FKT}, is the invariance of the integrability conditions (\ref{fourth}) under  projective transformations of the form
\begin{equation}
\tilde a=\frac{l_1(a, b, c)}{l(a, b, c)}, ~~ \tilde b=\frac{l_2(a, b, c)}{l(a, b, c)}, ~~\tilde c=\frac{l_3(a, b, c)}{l(a, b, c)}, ~~ \tilde f=\frac{f}{l(a, b, c)};
\label{proj}
\end{equation}
here $l, l_1, l_2, l_3$ are arbitrary (inhomogeneous) linear forms in $a, b, c$.  Introducing the quartic form
$$
F=Hd^4f-d^3fdH- 3 det (dM),
$$
one can verify that $\tilde F=l^4 F$, which establishes the $SL(4,
\R)$-invariance of the integrability conditions (\ref{fourth}).
Combined with obvious symmetries of the form
\begin{equation}
\tilde f= s f+\alpha a+ \beta b+ \gamma c+\delta,
\label{f}
\end{equation}
this provides a $20$-dimensional symmetry group of the problem.

\noindent {\bf Remark.} The class of  Euler-Lagrange equations
(\ref{Lag}) is form-invariant under a point group  generated by
arbitrary linear transformations  of the variables $x, y, t$ and
$u$. Obviously, point transformations preserve the integrability.
Since the prolongation of these transformations to the variables
$a,b,c$ and $f$ is given by (\ref{proj}), this explains the $SL(4,
\R)$-invariance of the integrability conditions (\ref{fourth}).

The main result of this section is the following

{\bf Theorem 2.} {\it The action of the symmetry group on the
$20$-dimensional moduli space of integrable Lagrangians possesses an
open orbit.}

\medskip

\centerline {\bf Proof:}

\medskip
\noindent The infinitesimal generators of the symmetry group (\ref{proj}), (\ref{f}) are the following vector fields:

\noindent 3 translations in $a, b, c$:
$$
 \frac{\partial}{\partial a}, ~~~ \frac{\partial}{\partial b}, ~~~ \frac{\partial}{\partial c};
 $$
 \noindent 9 linear transformations of $a, b, c$:
 $$
a \frac{\partial}{\partial a}, ~~~ b\frac{\partial}{\partial a}, ~~~ c\frac{\partial}{\partial a}, ~~~ a\frac{\partial}{\partial b}, ~~~ b\frac{\partial}{\partial b}, ~~~ c\frac{\partial}{\partial b}, ~~~ a\frac{\partial}{\partial c}, ~~~ b\frac{\partial}{\partial c}, ~~~ c\frac{\partial}{\partial c};
$$
\noindent 3 projective transformations of $a, b, c, f$:
$$
a^2\frac{\partial}{\partial a}+ab\frac{\partial}{\partial b}+ac\frac{\partial}{\partial c}+af\frac{\partial}{\partial f}, ~~~
ab\frac{\partial}{\partial a}+b^2\frac{\partial}{\partial b}+bc\frac{\partial}{\partial c}+bf\frac{\partial}{\partial f}, ~~~
ac\frac{\partial}{\partial a}+bc\frac{\partial}{\partial b}+c^2\frac{\partial}{\partial c}+cf\frac{\partial}{\partial f};
$$
\noindent moreover, we have 5 extra generators corresponding to the transformations (\ref{f}):
$$
 \frac{\partial}{\partial f}, ~~~ a\frac{\partial}{\partial f}, ~~~ b\frac{\partial}{\partial f}, ~~~ c\frac{\partial}{\partial f}, ~~~ f\frac{\partial}{\partial f}.
 $$
The main idea of the proof is to prolong these infinitesimal generators  to the $20$-dimensional moduli space  of solutions of the involutive system (\ref{fourth}). We point out that, since all fourth order derivatives of $f$ are explicitly known, this moduli space can be identified with the values of $f$ and its partial derivatives $f_i$, $f_{ij}$,  $f_{ijk}$ up to order three ($20$ parameters altogether).
The prolongation can be calculated as follows:

\noindent (1) Following the standard notation adopted in the symmetry analysis of differential equations
\cite{Ibragimov, Olver}, we introduce the variables $x^1=a, ~  x^2=b, ~ x^3=c$
and represent each of the above generators in the form
$$
 \xi^i\frac{\partial}{\partial x^i}+\eta \frac{\partial}{\partial f};
$$
here $\xi^i$ and $\eta$ are  functions of $x^i$ and $f$.

\noindent (2) Prolong   infinitesimal generators to the third order jet space with coordinates $x^i, f, f_i, f_{ij}, f_{ijk}$,
$$
 \xi^i\frac{\partial}{\partial x^i}+\eta \frac{\partial}{\partial f}+
 \zeta_i\frac{\partial}{\partial f_i}+
 \zeta_{ij}\frac{\partial}{\partial f_{ij}}+ \zeta_{ijk}\frac{\partial}{\partial f_{ijk}},
$$
where $\zeta_i$, $\zeta_{ij}$ and $\zeta_{ijk}$ are calculated according to the standard prolongation formulae
\begin{equation}
\zeta_i=D_i(\eta)-f_sD_i(\xi^s), ~~~ \zeta_{ij}=D_j{\zeta_i}-f_{is}D_j(\xi^s), ~~~  \zeta_{ijk}=D_k{\zeta_{ij}}-f_{ijs}D_k(\xi^s);
\label{prolong}
\end{equation}
here $D_i$ denotes the operator of total differentiation with respect to $x^i$.

\noindent (3) To eliminate the $\frac{\partial}{\partial x^i}$-terms we
subtract  the linear combination of total derivatives
$ \xi^i D_i$ from the prolonged operators where,  in $D_i$, it is sufficient to keep only the following terms:
$$
D_i=\frac{\partial}{\partial x^i}+f_i \frac{\partial}{\partial f}+f_{ij}\frac{\partial}{\partial f_j}+f_{ijk}\frac{\partial}{\partial f_{jk}}+f_{ijkl}\frac{\partial}{\partial f_{jkl}};
$$
notice that, since $f_{ijkl}$ are explicit functions of lower order derivatives, the resulting operators will be
well-defined vector fields on the $20$-dimensional moduli space with coordinates $f, f_i, f_{ij}, f_{ijk}$. Although these operators  will depend on the variables $x^i$ as on  parameters (indeed, the isomorphism of the moduli space with the space $f, f_i, f_{ij}, f_{ijk}$ depends on the choice of a point in the $x$-space), all algebraic properties of these operators will be $x$-independent.

\noindent (4) Finally, the dimension of the maximal orbit  equals the rank of the $20\times 20$ matrix  of  coefficients of these operators. It remains to point out that this rank equals $20$ for any `random' numerical choice of the values for
$x^i, f, f_i, f_{ij}, f_{ijk}$.

\section{Lagrangian densities in terms of modular forms}

In this section we provide explicit formulae for integrable Lagrangians in terms of modular forms. We  start  with the case $f(u_x, u_y, u_t)=u_xu_yg(u_t)$ (Sect. 3.1), where $g$ is shown to be an Eisenstein series $E_{1, 3}$: a special modular form of weight 1 and level three. The case $f(u_x, u_y, u_t)=g(u_x, u_y)u_t$ and the general case $f(u_x, u_y, u_t)$ are discussed in Sect. 3.2 and 3.3, respectively. We demonstrate how the knowledge of the symmetry group of the problem allows one to linearize the complicated nonlinear equations for Lagrangian densities resulting from the  integrablity conditions.

\subsection{Lagrangian densities of the form $f=u_xu_yg(u_t)$}

The integrability conditions (\ref{fourth}) imply a single fourth order ODE for $g(z)$,
\begin{equation}
g''''(g^2g''-2g(g')^2)-9(g')^2(g'')^2+2gg'g''g'''+8(g')^3g'''-g^2(g''')^2=0;
\label{g}
\end{equation}
to comply with the standard notation,  the argument of $g$ is now
denoted by $z$. This equation enjoys a remarkable $SL(2,
\R)$-invariance inherited from (\ref{proj}):
\begin{equation}
\tilde z=\frac{\alpha z+ \beta }{\gamma z+ \delta}, ~~~ \tilde g= (\gamma z+ \delta) g;
\label{gsym}
\end{equation}
here $\alpha, \beta, \gamma, \delta$ are arbitrary constants such that $\alpha \delta -\beta \gamma=1$. Moreover, there is an obvious scaling symmetry $g\to \lambda g$. The equation (\ref{g}) can be linearized as follows. Introducing $h=g'/g$, we first rewrite it in  the form
\begin{equation}
h'''(h'-h^2)=h^6-3h^4h'+9h^2(h')^2-3(h')^3-4h^3h''+(h'')^2;
\label{h}
\end{equation}
the corresponding symmetry group modifies to
\begin{equation}
\tilde z=\frac{\alpha z+ \beta }{\gamma z+ \delta}, ~~~ \tilde h= (\gamma z+ \delta)^2 h+\gamma(\gamma z+ \delta).
\label{hsym}
\end{equation}
We point out that the same symmetry  occurs in the case of the Chazy
equation (\cite{A1}, p. 342), as well as its analogue discussed
recently in \cite{A2}. The presence of the $SL(2, \R)$-symmetry of
this type implies the linearizability of the equation under study.
One can formulate the following general statement which is, in fact,
contained in  \cite{Clarkson}:

{\bf Theorem 3.}  {\it Any  third order ODE of the form $F(z, h, h',
h'', h''')=0$, which is invariant under the action of $SL(2, \R)$ as
specified by (\ref{hsym}), can be linearized by a substitution
\begin{equation}
z=\frac{w_1}{w_2}, ~~~ h=\frac{d}{dz}\ln w_2
\label{sub}
\end{equation}
where $w_1(t)$ and $w_2(t)$ are two linearly independent solutions of a linear equation $d^2w/dt^2=V(t)  w$ with the Wronskian $W$ normalized as  $W=w_2dw_1/dt-w_1dw_2/dt=1$ (the potential $V(t)$ depends on the given third order ODE, and can be effectively reconstructed).

In particular, the general solution of the equation (\ref{h}) is given by parametric formulae (\ref{sub})
where $w_1(t)$ and $w_2(t)$ are two linearly independent solutions of the linear equation $d^2w/dt^2=\frac{2}{9}(\cosh^{-2}t) w$ with $W=1$.}

\medskip

\centerline {\bf Proof:}

\medskip

\noindent  To establish the first part of the theorem we  essentially reproduce the calculation from Sect. 5 in \cite{Clarkson}. Let us consider a linear ODE $d^2w/dt^2=V(t)  w$,
 take two linearly independent solutions
$w_1(t)$, $w_2(t)$ with $W=1$, and introduce the new dependent and independent variables  $h, z$ by parametric relations
$$
z=\frac{w_1}{w_2}, ~~~ h=\frac{d}{dz}\ln w_2.
$$
Using the readily verifiable formulae $dt/dz=w_2^2$ and $ h=w_2dw_2/dt$, one obtains the identities
$$
h'-h^2=w_2^4\  V,
$$
$$
h''-6hh'+4h^3=w_2^6\ dV/dt
$$
and
$$
h'''-12hh''-6(h')^2+48h^2h'-24h^4=w_2^8\ d^2V/dt^2;
$$
here prime denotes differentiation with respect to $z$.  Thus, one arrives at the relations
$$
\begin{array}{c}
I_1=\frac{(h''-6hh'+4h^3)^2}{(h'-h^2)^3}=\frac{(dV/dt)^2}{V^3},   \\
\ \\
I_2=\frac{h'''-12hh''-6(h')^2+48h^2h'-24h^4}{(h'-h^2)^2}=\frac{d^2V/dt^2}{V^2}. \\
\end{array}
$$
We point out that $I_1$ and $I_2$ are the simplest second- and third-order differential invariants of the action (\ref{hsym}) whose infinitesimal generators, prolonged to the third jets $z, h, h', h'', h'''$, are of the form
$$
\begin{array}{c}
X_1=\partial_z, ~~~ X_2=z\partial_z-h\partial_h-2h'\partial_{h'}-3h''\partial_{h''}-4h'''\partial_{h'''}, \\
\ \\
X_3=z^2\partial_z-(2zh+1)\partial_h-(2h+4zh')\partial_{h'}-(6h'+6zh'')\partial_{h''}-(12h''+8zh''')\partial_{h'''};
\end{array}
$$
notice the standard commutation relations $[X_1,  X_2]=X_1, \  [X_1,
X_3]=2X_2, \ [X_2,  X_3]=X_3$. One can verify that the Lie
derivatives of $I_1, I_2$ with respect to $X_1, X_2, X_3$ are indeed
zero. Thus, any third order ODE which is invariant under the $SL(2,
\R)$-action (\ref{hsym}), can be represented in the form
$I_2=F(I_1)$ where $F$ is an arbitrary function of one variable. The
corresponding potential $V(t)$ has to satisfy the relation
$\frac{d^2V/dt^2}{V^2}=F\left(\frac{(dV/dt)^2}{V^3} \right).$

This simple scheme produces some of the well-known equations, for instance,  the relation $I_2=-24$  implies the Chazy equation for $h$, that is,
$h'''-12hh''+18(h')^2=0$. The corresponding potential satisfies the equation $d^2V/dt^2=-24V^2$.

Similarly, the choice $I_2=I_1-8$ results in the ODE $h'''=4hh''-2(h')^2+\frac{(h''-2hh')^2}{h'-h^2}$ which, under the substitution $h=y/2$, coincides with the equation (4.7) from \cite{A2}.
The potential $V$ satisfies the equation $Vd^2V/dt^2=(dV/dt)^2-8V^3$.

Finally, the relation $I_2=I_1-9$ coincides with (\ref{h}). The corresponding potential $V$ satisfies the equation $Vd^2V/dt^2=(dV/dt)^2-9V^3$. It remains to point out that, up to elementary equivalence transformations, the general solution of the last  equation for $V$ is given by $V=\frac{2}{9}\cosh^{-2}t $.

\medskip

Since the equation $d^2w/dt^2=\frac{2}{9}\cosh^{-2}t \ w$ is related to the hypergeometric equation
$s(1-s)w_{ss}+(1-2s)w_s-\frac{2}{9}w=0$, corresponding to the parameter values $a =1/3, \ b =2/3, \ c=1$, by a change of variables $s/(1-s)=e^{2t}$,  we can reformulate the above Theorem as follows:

\medskip

\noindent{\bf Proposition 1.} {\it The general solution of the equation (\ref{h}) is given by parametric formulae (\ref{sub})
where $w_1(s)$ and $w_2(s)$ are two linearly independent solutions of the hypergeometric equation $s(1-s)w_{ss}+(1-2s)w_s-\frac{2}{9}w=0$ with the Wronskian normalized as  $w_2dw_1/ds-w_1dw_2/ds=1/(2s(1-s))$.}

\medskip

As  $h=g'/g$, this immediately implies the following formula for the general solution of (\ref{g}):

\noindent{\bf Proposition 2.} {\it The general solution of the equation (\ref{g}) is given by parametric formulae
$$
z=\frac{w_1}{w_2}, ~~~ g= w_2,
$$
where $w_1(s)$ and $w_2(s)$ are two linearly independent solutions to the hypergeometric equation $s(1-s)w_{ss}+(1-2s)w_s-\frac{2}{9}w=0$ with the Wronskian normalized as  $w_2dw_1/ds-w_1dw_2/ds=1/(2s(1-s))$.}

\medskip


One can construct the following explicit solution of the equation (\ref{g}),
\begin{equation}
g(z)=\sum_{(\alpha,\beta)\in\Z^2}q^{(\alpha^2-\alpha\beta+\beta^2)}=1+6q+6q^3+6q^4+12q^7+....;
\label{g1}
\end{equation}
here $q=e^{2\pi i z}$. To get a real-valued solution, one has to restrict $z$ to the imaginary axis. This function is known as the Eisenstein series $E_{1, 3}(z)$. Equivalently, it can be defined by the formula
$$
g(z)=E_{1, 3}(z)=1+6\sum_{n=1}^{\infty}\left(\sum_{d\vert n}\chi_3(d)\right) q^n
$$
where $\chi_3$ denotes the Legendre symbol mod $3$ (that is, $\chi_3(d)=0$ if $d\equiv 0 $ mod $3$,
$\chi_3(d)=1$ if $d\equiv 1 $ mod $3$, and $\chi_3(d)=-1$ if $d\equiv 2 $ mod $3$). The Eisenstein series transforms as $g(\frac{\alpha z+\beta}{\gamma z+\delta})=\chi_3(\delta) (\gamma z+\delta)g(z)$ under the Hecke congruence subgroup $\Gamma_0(3)$ defined as
$$
\left(\begin{array}{cc}
\alpha & \beta \\
\gamma  & \delta
\end{array}
\right)\in \Gamma_0(3) \subset SL(2, \Z) ~~~ {\rm if} ~~~
\left(\begin{array}{cc}
\alpha &\beta \\
\gamma & \delta
\end{array}
\right)\equiv\left(\begin{array}{cc}
\alpha&\beta \\
0 & \delta
\end{array}
\right)  ~ {\rm mod} ~ 3.
$$
It follows that $g(z)$ is a modular form of weight one and level $3$,  namely,
$g(\frac{\alpha z+\beta}{\gamma z+\delta})=(\gamma z+\delta)g(z)$ where
$$
\left(\begin{array}{cc}
\alpha & \beta \\
\gamma  & \delta
\end{array}
\right)\in SL(2, \Z) ~~~ {\rm and} ~~~ \left(\begin{array}{cc}
\alpha &\beta \\
\gamma & \delta
\end{array}
\right)\equiv\left(\begin{array}{cc}
1&\beta \\
0 & 1
\end{array}
\right)  ~ {\rm mod} ~ 3.
$$
The function $g(z)$ can also be written in the
form involving summation over $\N$ only,
$$
g(z)=1-6\sum_{k\in\N}\left(\frac{q^{3k-1}}{1-q^{3k-1}}-\frac{q^{3k-2}}{1-q^{3k-2}}\right).
$$

 {\bf Theorem 4.} {\it The function $g(z)$ is a solution of the
differential equation (\ref{g}).}

\medskip

\noindent \centerline{\bf Proof:}

\medskip

\noindent Recall that, given a modular form $g(z)$ of weight $k$, its Rankin-Cohen brackets
$[g, g]_2$ and $[g, g]_4$ are defined as follows:
$$
[g, g]_2=(k+1)\left(kgg''-(k+1)(g')^2\right),
$$
and
$$
[g, g]_4=(k+2)(k+3)\left(\frac{k(k+1)}{12} gg''''-\frac{(k+1)(k+3)}{3}g'g'''+\frac{(k+2)(k+3)}{4}(g'')^2\right);
$$
these are known to be modular forms of   weights $2k+4$ and $2k+8$, respectively (we use the normalization of \cite{Zagier}).
In our case $k=1$, so that we get
$$
G=[g, g]_2=2(gg''-2(g')^2), ~~~
[g, g]_4=2(gg''''-16g'g'''+18(g'')^2),
$$
weights $6$ and $10$, respectively. One can verify that, up to a constant multiple,  the left hand side of the equation (\ref{g}) can be represented in the form
$$
7[g, g]_4[g, g]_2+[G, G]_2,
$$
which shows that it is a modular form (in fact, a cusp form) of weight $16$ with respect to the
same group. To show that this form vanishes identically we recall that the dimension of the space of cusp forms of weight 16 and level 3 equals 4, and the order of zero cannot exceed $5$. Thus, it is sufficient to verify the vanishing of the first five coefficients in the decomposition of this form as a power series in $q=\exp(2\pi i z)$. This can be done by a direct calculation.

{\bf Remark.} The relation between the modular form (\ref{g1}) and the hypergeometric equation from  the Proposition 2 can be summarized as follows. Choosing a basis of solutions of the hypergeometric equation in the form
$$
w_2=1+\frac{2}{9} s+\frac{10}{81}s^2+...,  ~~~ w_1=w_2\ln s+\frac{5}{9}s+\frac{57}{162}s^2+...,
$$
one obtains parametric equations
$$
z=\frac{w_1}{w_2}=\ln s+\frac{5}{9}s+\frac{37}{162}s^2+..., ~~~ g=w_2=1+\frac{2}{9} s+\frac{10}{81}s^2+....
$$
Solving the first equation for $s$ in the form  $s=e^z+ae^{2z}+be^{3z}+...$ one arrives at
$s= e^z-\frac{5}{9}e^{2z}+\frac{19}{81}e^{3z}+...$. The substitution into the second equation implies the expression for $g(z)$ in the form $g(z)=1+pe^z+qe^{3z}+...$ which, up to an appropriate affine transformation of $z$, coincides with (\ref{g1}).

\subsection{Lagrangian densities of the form $f=g(u_x,  u_y)u_t$}

As follows from \cite{FKT}, the integrability conditions  (\ref{fourth}) result in a system of five equations expressing all fourth order
partial derivatives of $g(a, b)$ in terms of its lower order derivatives. In symbolic form, this system can be represented as
follows:
\begin{equation}
d^4g=d^3g\frac{dh}{h}+6\frac{dg}{h}{ det} (dm)+3\frac{(dg)^2}{h}det (dn).
\label{fourth1}
\end{equation}
Here  $d^sg$  are  symmetric differentials of $g$,  the matrices $m$ and $n$ are defined as
$$
m=\left(\begin{array}{ccc}
0 & g_{a} & g_{b}\\
g_{a} & g_{aa} & g_{ab}  \\
g_{b} & g_{ab} & g_{bb}
\end{array}
\right), ~~~~
n=\left(\begin{array}{cc}
g_{aa} & g_{ab}  \\
g_{ab} & g_{bb}
\end{array}
\right),
$$
and
$$
h=-det (m)=g_{b}^2g_{aa}-2g_{a}g_{b}g_{ab}+g_{a}^2g_{bb}.
$$
The non-degeneracy of the Lagrangian density $f(a, b, c)=g(a, b) c$   is equivalent to the condition $h\ne 0$. One can show that
the over-determined
system (\ref{fourth1})  is in involution, and its solution
space is $10$-dimensional (indeed, the values of partial derivatives of $g$ up to
order 3 at a  point $(a_0, b_0)$  amount to  $10$ arbitrary constants). The system
(\ref{fourth1}) is invariant under a $10$-dimensional group of Lie-point symmetries which consists of arbitrary projective
transformations of $a$ and $ b$, isomorphic to $SL(3, \R)$, along with transformations of the form
$g\to \alpha g+\beta, \ \alpha, \beta ={\rm const}$. The corresponding infinitesimal generators include

\begin{equation}
\begin{array}{c}
\noindent {\rm 2~translations}: ~~
\displaystyle  \frac{\partial}{\partial a}, ~~~ \frac{\partial}{\partial b};\\
\ \\
 \noindent {\rm 4~ linear~transformations}:
\displaystyle ~~~ a \frac{\partial}{\partial a}, ~~~ b\frac{\partial}{\partial a},  ~~~ a\frac{\partial}{\partial b}, ~~~ b\frac{\partial}{\partial b}; \\
\ \\
\noindent {\rm 2~projective~transformations}:
~~~
\displaystyle a^2\frac{\partial}{\partial a}+ab\frac{\partial}{\partial b}, ~~~
ab\frac{\partial}{\partial a}+b^2\frac{\partial}{\partial b};\\
\ \\
\noindent {\rm 2~affine~transformations~of} ~ g:
~~~
\displaystyle \frac{\partial}{\partial g}, ~~~ g\frac{\partial}{\partial g}.
 \end{array}
 \label{sym2}
 \end{equation}
We will demonstrate that the existence of this symmetry group allows one to linearize the integrability conditions (\ref{fourth1}).
For this purpose we consider a linear system of the form
\begin{eqnarray}
z_{xx}&=&Mz_x-Iz_y+A z, \nonumber \\
z_{xy}&=&-Nz_x-Mz_y+B z,  \label{lin} \\
z_{yy}&=&-Jz_x+Nz_y+C z, \nonumber
\end{eqnarray}
where the coefficients $I, J, M, N, A, B, C$ are certain functions of $x, y$ which have to satisfy the compatibility conditions resulting from the requirement of consistency of the equations (\ref{lin}):
$$
A=2(M^2+IN)+I_y-M_x, ~~~ B=M_y+N_x+IJ-MN, ~~~ C= 2(N^2+JM)+J_x-N_y,
$$
along with two extra relations involving $I, J, M, N$ only:
$$
\begin{array}{c}
J_{xx}=2N_{xy}+M_{yy}-3(N^2+JM)_x-3NM_y+2JI_y+IJ_y, \\
I_{yy}=2M_{xy}+N_{xx}-3(M^2+IN)_y-3MN_x+2IJ_x+JI_x.
\end{array}
$$
These relations imply that the space of solutions of the linear system (\ref{lin}) is three-dimensional. Notice that any involutive second order linear system of the form $z_{ij}=\Gamma^k_{ij}z_k+g_{ij}z$ with two independent variables $x, y$
can be reduced to the form (\ref{lin}) by a gauge transformation $z\to \varphi(x, y) z$. This is a standard normalization in the
theory of multi-dimensional Schwarzian derivatives \cite{Sasaki}. It implies that the Wronskian $W$ of any three linearly
independent solutions,
$$
W=det \left[
\begin{array}{ccc}
z_1 & z_2 &z_3 \\
(z_1)_x & (z_2)_x & (z_3)_x \\
(z_1)_y & (z_2)_y & (z_3)_y
\end{array}
\right],
$$
is  constant: $W_x=W_y=0$. Let us choose  three linearly independent solutions $z_i(x, y)$ with the Wronskian normalized as $W=1$, and introduce the new independent variables
$$
a=\frac{z_1}{z_3}, ~~~ b=\frac{z_2}{z_3}.
$$
Let us consider $x$ and $y$ as functions of $a, b$:
$$
x=x(a, b), ~~~ y=y(a, b).
$$
A direct calculation shows that these functions satisfy the following nonlinear system:
\begin{equation}
\begin{array}{c}
x_a^2x_{bb}-2x_ax_bx_{ab}+x_b^2x_{aa}=(x_ay_b-y_ax_b)^2J, \\
\ \\
y_a^2y_{bb}-2y_ay_by_{ab}+y_b^2y_{aa}=(x_ay_b-y_ax_b)^2I, \\
\ \\
y_ax_{aa}-x_ay_{aa}=3Nx_ay_a^2+Jy_a^3-Ix_a^3-3Mx_a^2y_a, \\
\ \\
x_by_{bb}-y_bx_{bb}=3Mx_b^2y_b+Ix_b^3-Jy_b^3-3Nx_by_b^2;
\end{array}
\label{xy}
\end{equation}
here $I, J, M, N$ are the same functions of $x, y$ as in (\ref{lin}).  The  system (\ref{xy}) is in involution,  and its solution space, which  is $8$-dimensional,  possesses a transitive action of $SL(3, \R)$: indeed, there is an $SL(3, \R)$-freedom in the choice of a basis $z_1, z_2, z_3$.
Conversely, one can show that any involutive system of four second order PDEs for two functions
$x(a, b)$ and $y(a, b)$ which is invariant under a transitive projective  action of $SL(3, \R)$,  has the form (\ref{xy}), and comes from a linear system (\ref{lin}). This immediately follows from the  equivalent representation of the system (\ref{xy}),
\begin{equation}
\begin{array}{c}
\displaystyle \frac{x_a^2x_{bb}-2x_ax_bx_{ab}+x_b^2x_{aa}}{(x_ay_b-y_ax_b)^2}=J(x, y), \\
\ \\
\displaystyle \frac{y_a^2y_{bb}-2y_ay_by_{ab}+y_b^2y_{aa}}{(x_ay_b-y_ax_b)^2}=I(x, y), \\
\ \\
\displaystyle \frac{y_b^2x_{aa}+y_a^2x_{bb}+2x_by_by_{aa}+2x_ay_ay_{bb}-2y_ay_bx_{ab}-2(x_ay_b+x_by_a)y_{ab}}{(x_ay_b-y_ax_b)^2}=-3M(x, y), \\
\ \\
\displaystyle \frac{x_a^2y_{bb}+x_b^2y_{aa}+2x_ay_ax_{bb}+2x_by_bx_{aa}-2x_ax_by_{ab}-2(x_ay_b+x_by_a)x_{ab}}{(x_ay_b-y_ax_b)^2}=-3N(x, y).
\end{array}
\label{xy1}
\end{equation}
The variables $x, y$ and the left hand sides of (\ref{xy1}) form a basis of differential invariants of the  $SL(3, \R)$-action  extended to  second order jet space with coordinates $a, b, x, y, x_a, x_b, y_a, y_b$,  $x_{aa}, x_{ab}, x_{bb}, y_{aa}, y_{ab}, y_{bb}$.  Thus, any system with the required symmetry
properties can be obtained by expressing the four second order differential invariants as functions of $x$ and $ y$. The expressions in the left hand sides of (\ref{xy1}) are related to the two-dimensional Schwarzian derivatives \cite{Sasaki}.

Let us return to the integrability conditions (\ref{fourth1}). Extending the action of the symmetry generators (\ref{sym2}) to the third order jet space with coordinates $a, b, g, g_a, g_b, g_{aa}, g_{ab}, g_{bb}$,  $g_{aaa}, g_{aab}, g_{abb}, g_{bbb}$ according to the prolongation formulae (\ref{prolong}), one obtains 10 vector fields on a 12-dimensional space; thus, there exist two differential invariants, which we will denote by $x$ and $y$, respectively (the explicit formulae for $x$ and $y$ in terms of $g$ and its derivatives are provided below). We claim that $x$ and $y$, viewed as functions of $a$ and $b$, satisfy a system of the form (\ref{xy})
(equivalently, (\ref{xy1})). Indeed, the action of the $SL(3, \R)$ part of the symmetry group on the variables $a, b, x, y$ is exactly the same as for the system  (\ref{xy}). The passage from $a, b, g$ to
$a, b, x, y$ can be viewed as a factorization of the integrability equations (\ref{fourth1}) by the two-dimensional affine  group corresponding to the last two generators (\ref{sym2}), which respects the action of $SL(3, \R)$. To write down the expressions for the differential invariants $x$ and $y$ we introduce the following notation:
$$
\displaystyle Z^1=\frac{\displaystyle \frac{g_{aaa}}{g_a}-\frac{3}{2}\frac{g_{aa}^2}{g_a^2}}{(g_a^2g_{bb}-2g_ag_bg_{ab}+g_b^2g_{aa})^2}\ g_a^2g_b^4, ~~~~~
\displaystyle Z^2=\frac{\displaystyle \frac{g_{bbb}}{g_b}-\frac{3}{2}\frac{g_{bb}^2}{g_b^2}}{(g_a^2g_{bb}-2g_ag_bg_{ab}+g_b^2g_{aa})^2}\ g_a^4g_b^2,
$$
\bigskip
$$
\displaystyle V^1=\frac{\displaystyle \frac{g_{aab}}{g_a}-2\frac{g_{aa}g_{ab}}{g_a^2}+\frac{1}{2}\frac{g_b}{g_a^3}g_{aa}^2}{(g_a^2g_{bb}-2g_ag_bg_{ab}+g_b^2g_{aa})^2}\ g_a^3g_b^3, ~~~~~
\displaystyle V^2=\frac{\displaystyle \frac{g_{abb}}{g_b}-2\frac{g_{bb}g_{ab}}{g_b^2}+\frac{1}{2}\frac{g_a}{g_b^3}g_{bb}^2}{(g_a^2g_{bb}-2g_ag_bg_{ab}+g_b^2g_{aa})^2}\ g_a^3g_b^3.
$$
Moreover, let
$$
\Sigma=Z^1-3V^1-(Z^2-3V^2), ~~~ S=Z^1+Z^2-V^1-V^2.
$$
Then the invariants $x$ and $y$ can be chosen in the following form:
$$
x=\frac{[\Sigma^3+9\Sigma S+18(Z^1-Z^2)]^2}{[\Sigma^2+6S+3]^3},
~~~~
y=\frac{\Sigma^2-S^2+4\Sigma(V^1-V^2)+4(V^1+V^2)-1}{[\Sigma^2+6S+3]^2};
$$
notice that the expressions for $x$ and $y$ are manifestly symmetric under the interchange of indices $1\leftrightarrow 2$. The functions $x(a, b)$ and $y(a, b)$ satisfy a system of the form (\ref{xy}) which is invariant under the action of $SL(3, \R)$ as specified above. The explicit formulae for the coefficients $I, J, M, N$, as well as the properties of the corresponding linear system (\ref{lin}), will be discussed elsewhere.

\subsection{General case: Lagrangian densities of the form $f(u_x, u_y, u_t)$}

Let us begin by introducing  special functions which will appear in
the general formula for $f$.  The first one is  a theta-function of
order one \cite{mamford} (with modular parameter
$\tau=\varepsilon$),
\begin{equation}
\label{thetaser}\theta(z)=\sum_{k\in\Z}(-1)^k\exp\left(2\pi
i(kz+\frac{k(k-1)}{2}\varepsilon)\right),
\end{equation}
which is known to satisfy the relations
$$\theta(z+1)=\theta(z),~~~ \theta(z+\varepsilon)=-\exp(-2\pi
iz)\theta(z);
$$
here  $\varepsilon=\frac{1}{2}(1+\sqrt{3}i)$. Note that
$\varepsilon$ is a primitive 6th root of unity. In particular,
$\varepsilon^3=-1$ and $\varepsilon^2=\varepsilon-1$. It is known
that $\theta(0)=0$, and this is the only zero of the function
$\theta(z)$ modulo 1, $\varepsilon$. Moreover,
$\theta^{\prime}(0)\ne0$. Next we define a function $\tilde
\theta(z)$ which differs from $\theta(z)$ by an exponential factor:
\begin{equation}\label{thetanorm}\tilde{\theta}(z)=\frac{1}{\theta^{\prime}(0)}\exp\left(-2\pi
i(\frac{\varepsilon}{3}-\frac{1}{6})z^2-\pi
iz\right)\theta(z).\end{equation} It can be readily verified that
\begin{equation}
\label{theta}
\begin{array}{c}
\tilde{\theta}(z+1)=
\exp(-\frac{2\pi
i}{3}((2\varepsilon-1)z+\varepsilon-2))\tilde{\theta}(z), \\
\ \\
\tilde{\theta}(z+\varepsilon)= \exp(-\frac{2\pi
i}{3}((\varepsilon+1)z+\varepsilon-2)))\tilde{\theta}(z).
\end{array}
\end{equation}
These relations  imply
\begin{equation}\label{thetag}
\tilde{\theta}(z+\alpha+\beta\varepsilon)=\exp\left(-2\pi
i((\frac{2\varepsilon-1}{3}\alpha+\frac{\varepsilon+1}{3}\beta)z+
\frac{\varepsilon+1}{3}\alpha^2+\frac{\varepsilon+1}{3}\alpha\beta+\frac{\varepsilon+1}{3}\beta^2)\right)\tilde{\theta}(z)
\end{equation}
where $\alpha,\beta\in\Z$. One can also show that
$$
\tilde{\theta}(\varepsilon
z)=\varepsilon\tilde{\theta}(z),
$$
which implies
\begin{equation}
\tilde{\theta}(z)=\sum_{j\geq0}a_jz^{6j+1};
\label{series}
\end{equation}
note that $a_0=\tilde{\theta}^{\prime}(0)=1$.

Finally, let us define the `master-density'  $f$ by the
formula
\begin{equation}
\label{lagr}
f(x,y,z)=xy+\sum_{(k,l)\in\Z^2\setminus0}\frac{\tilde{\theta}((k-\varepsilon
l)x)\tilde{\theta}((k-\varepsilon l)y)}{(k-\varepsilon
l)^2}\exp\left(\frac{2\pi i}{3}(k^2-kl+l^2)z\right),
\end{equation}
which we claim to be a `generic' solution of (\ref{fourth})\footnote{We did not prove that $f$ satisfies the system
(\ref{fourth}), although computer calculations support this claim.}.  For convenience, the arguments of $f$ are now denoted by
$x, y, z$.
 From  (\ref{thetag}) one can derive the following modular properties of $f$:
$$f(x+1, y, z+(2\varepsilon-1)x+\varepsilon+1)=f(x,y,z)+y,$$
\begin{equation}
\label{mod}
f(x+\varepsilon,y,z+(\varepsilon+1)x+\varepsilon+1)=f(x,y,z)+\varepsilon
y,
\end{equation}
$$
f(x,y+1,z+(2\varepsilon-1)y+\varepsilon+1)=f(x,y,z)+x,$$$$
f(x,y+\varepsilon,z+(\varepsilon+1)y+\varepsilon+1)=f(x,y,z)+\varepsilon
x.
$$
Substituting (\ref{series}) into (\ref{lagr}) we obtain an alternative representation for $f$,
$$
f(x,y,z)=xy+\sum_{(k,l)\in\Z^2\setminus0, \ m,n\geq0}a_ma_nx^{6m+1}y^{6n+1}\left((k-\varepsilon
l)^{6(m+n)}\exp\left(\frac{2\pi
i}{3}(k^2-kl+l^2)z\right)\right)=
$$
$$
\sum_{m,n\geq0}a_ma_nx^{6m+1}y^{6n+1}g_{m+n}(z)
$$
where
$$
g_n(z)=\sum_{k,l\in\Z^2}\frac{1}{2}((k-\varepsilon
l)^{6n}+(\varepsilon k- l)^{6n})\exp\left(\frac{2\pi
i}{3}(k^2-kl+l^2)z\right).
$$
It is known \cite{ogg} that $g_n(z)$ is a modular form, namely,
$g_n(\frac{\alpha z+\beta}{\gamma z+\delta})=(\gamma
z+\delta)^{6n+1}g_n(z)$ where $\left(\begin{array}{cc}
\alpha &\beta \\
\gamma & \delta
\end{array}
\right)\in SL(2, \Z)$ and $\left(\begin{array}{cc}
\alpha &\beta \\
\gamma & \delta
\end{array}
\right)\equiv\left(\begin{array}{cc}
1&0 \\
\gamma & 1
\end{array}
\right) $mod $3$. Indeed, coefficients at the exponents are harmonic
polynomials with respect to the quadratic form $k^2-kl+l^2$.
This gives us the following modular property of $f(x,y,z)$ with
respect to the same group:
\begin{equation}\label{mod1}
f\left(\frac{x}{\gamma z+\delta}, \ \frac{y}{\gamma z+\delta}, \ \frac{\alpha z+\beta}{\gamma z+\delta}\right)=f(x, y, z)\frac{1}{\gamma z+\delta}.
\end{equation}
Note
that $g(z)=g_0(z)$ is a solution of the differential equation
(\ref{g}) and is given by (\ref{g1}) (up to a rescaling of $z$). Functions $g_n(z)$ can be
represented as  rational differential functions in $g(z)$, for
example, $g_1(z)=g(z)^2g^{\prime\prime}(z)-2g(z)g^{\prime}(z)^2=\frac{1}{2}g [g, g]_2$ where $[g, g]_2$ is the Rankin-Cohen bracket (see the proof of Theorem 4).
Transformations (\ref{mod}) and (\ref{mod1}) generate a discrete
subgroup $\Gamma \subset SL(4, \C)$ which plays the role of the
modular group for  $f(x, y, z)$. The subgroup $\Gamma$ has the
following algebraic description. Let $\D\subset\C^3$ be a domain in
$\C^3$ defined by
$$\D=\{(x,y,z)\in\C^3;~~~|x|^2+|y|^2<\frac{2}{\sqrt 3}\Im z\}.$$
Note that $\D$ is a complex hyperbolic ball. One can check that the
series (\ref{lagr}) converges exactly in the domain $\D$. Let
$G\subset SL(4, \C)$ be a group defined by
$$G=\{A\in SL(4, \C);~~~AJA^*=J\}$$
where
$$
J=\left(\begin{array}{cccc}
1&0&0&0 \\0&1&0&0\\0&0&0&-\sqrt3i \\
0 & 0&\sqrt3i&0
\end{array}
\right),
$$
 and $A^*$ stands for the Hermitian conjugate  of
$A$. One can check that the complex hyperbolic ball $\D$ is an orbit
of $G$ under its standard projective action on $\C P^3$:  if $A=(a_{ij})\in G$ and
$(x,y,z)=(x:y:z:1)\in\C P^3$, then
$$
A(x,y,z)=\left(\frac{a_{11}x+a_{12}y+a_{13}z+a_{14}}{a_{41}x+a_{42}y+a_{43}z+a_{44}}, \
\frac{a_{21}x+a_{22}y+a_{23}z+a_{24}}{a_{41}x+a_{42}y+a_{43}z+a_{44}}, \
\frac{a_{31}x+a_{32}y+a_{33}z+a_{34}}{a_{41}x+a_{42}y+a_{43}z+a_{44}}\right).
$$
In our context, the domain $\D$ and the group $G$ play a role
similar to that of the upper half plane and its automorphism group $
SL(2, \R)$ in the classical theory of modular forms. Let
$\Gamma\subset G$ be a discrete subgroup of $G$ consisting of
matrices $A=(a_{ij})\in G$ with the following properties:
$a_{ij}\in\Z[\varepsilon]$ and
$$
A\equiv\left(\begin{array}{cccc}
1&0&a_{13}&a_{14} \\0&1&a_{23}&a_{24}\\0&0&1&0 \\
0 & 0&a_{43}&1
\end{array}
\right) \  {\rm mod }\  (1+\varepsilon).
$$
Here $\Z[\varepsilon]=\{m+\varepsilon n;~~~n,m\in\Z\}$, and $a\equiv
b$ mod$(1+\varepsilon)$ for $a,b\in\Z[\varepsilon]$ means
$\frac{a-b}{1+\varepsilon}\in\Z[\varepsilon]$. Note that all
matrices corresponding to the transformations (\ref{mod}) and
(\ref{mod1}) belong to $\Gamma$. We conjecture that the group
$\Gamma$ is generated by these transformations.

{\bf Remark 1: limiting cases.} Lagrangian densities of the form $xg(y,z)$ can be obtained as  $\lim_{t\to0}
\frac{f(tx,y,z)}{t}$
from $f(x,y,z)$ as defined by (\ref{lagr}). This gives  the function
$g(y,z)$ in the form
\begin{equation}
\label{lagr1}
g(y,z)=y+\sum_{(k,l)\in\Z^2\setminus0}\frac{\tilde{\theta}((k-\varepsilon
l)y)}{k-\varepsilon l}\exp\left(\frac{2\pi i}{3}(k^2-kl+l^2)z\right)=\sum_{n\geq0}a_ny^{6n+1}g_{n}(z).
\end{equation}
This function is defined on the domain
$\{(y,z)\in\C^2;~~~|y|^2<\frac{2}{\sqrt 3}\Im z\}$ and satisfies the following modular
properties:
$$g(y+1,z+(2\varepsilon-1)y+\varepsilon+1)=g(y,z)+1,$$
\begin{equation}
\label{mod2}g(y+\varepsilon,z+(\varepsilon+1)y+\varepsilon+1)=g(y,z)+\varepsilon,
\end{equation}
$$
g\left( \frac{y}{\gamma z+\delta}, \ \frac{\alpha z+\beta}{\gamma z+\delta}\right)=g(y, z).
$$
Similarly, Lagrangian densities of the form
$xyg(z)$ can be obtained as  $\lim_{t\to0} \frac{f(tx, ty, z)}{t^2}$. This brings us back to the modular form $g(z)$ discussed in Sect. 3.1.

{\bf Remark 2.} Computer experiments show that  solutions of
the system (\ref{fourth}) can also be sought in the form of  a power series,
$${f}(x,y,z)=\sum_{i,j,k\geq0}c_{ijk}x^{6i+1}y^{6j+1}z^{6k+1}.$$
Moreover,  $c_{ijk}=a_ia_ja_kb_{i+j+k}$ where $a_i$ are
the same as in (\ref{series}), and $b_i$ is yet another sequence of complex
numbers. Taking a limit we obtain
$${g}(y,z)=\sum_{j,k\geq0}a_ja_kb_{j+k}y^{6j+1}z^{6k+1}$$
for  densities of the form $f(x, y, z)=xg(y, z)$,  and
$${g}(z)=\sum_{k\geq0}a_kb_{k}z^{6k+1}$$
for densities of the form $f(x, y, z)=xyg(z)$. We point out that integrable densities of this type differ from
the solutions constructed above by appropriate transformations from the equivalence group.

{\bf Remark 3.} The action of the $20$-dimensional equivalence group on the $23$-dimensional space of third order jets of the function $f(a, b,c)$ possesses three differential invariants which we denote by $x, y, z$. Viewed as functions of $a, b, c$, these invariants satisfy a nonlinear system which possesses a transitive action of $SL(4, \R)$. This system can be linearised following the procedure outlined in Sect. 3.2. We plan to report the details elsewhere.

{\bf Remark 4.} There exist a number of examples   of integrable Lagrangian densities expressible in terms of  elementary functions. One can mention, e.g.,   the four polynomial Lagrangians classified in \cite{F1}:
$$
f=u_xu_yu_t, ~~~  f=u_x^2u_y+u_yu_t,  ~~~ f=u_x^3/3+u_y^2-u_xu_t
$$
and
$$
f=u_x^4+2u_x^2u_t-u_xu_y-u_t^2.
$$
It would be  interesting to explicitly demonstrate how these (and other) examples can be obtained as degenerations of the `master-Lagrangian' constructed in Section 3.2, and to describe singular orbits of lower dimensions.

\section*{Acknowledgements}

We thank R. Halburd, J.  Harnad,  M. Pavlov, V. Sokolov and A. Veselov  for  numerous
helpful discussions. The research of EVF was partially supported by the EPSRC grant  EP/D036178/1, the European Union through the FP6 Marie Curie RTN project {\em ENIGMA} (Contract number MRTN-CT-2004-5652), and the ESF programme MISGAM.

\end{document}